%% file: MAIN-arxiv.tex
\documentclass[acmsmall,screen]{acmart}
\AtBeginDocument{%
  }

\setcopyright{none}
\renewcommand\footnotetextcopyrightpermission[1]{}

\input{package}

\author{Binhang Qi}
\email{qibh@nus.edu.sg}
\affiliation{%
  \institution{National University of Singapore}
  \country{Singapore}
}

\author{Yun Lin}
\authornote{Corresponding author.}
\email{lin\_yun@sjtu.edu.cn}
\affiliation{%
  \institution{Shanghai Jiao Tong University}
  \country{China}
}

\author{Xinyi Weng}
\email{nanakusa@sjtu.edu.cn}
\affiliation{%
  \institution{Shanghai Jiao Tong University}
  \country{China}
}

\author{Yuhuan Huang}
\email{hyh0u0@sjtu.edu.cn}
\affiliation{%
  \institution{Shanghai Jiao Tong University}
  \country{China}
}

\author{Chenyan Liu}
\email{chenyan@u.nus.edu}
\affiliation{%
  \institution{National University of Singapore}
  \country{Singapore}
}

\author{Hailong Sun}
\email{sunhl@buaa.edu.cn}
\affiliation{%
  \institution{Beihang University}
  \country{China}
}

\author{Zhi Jin}
\email{zhijin@whu.edu.cn}
\affiliation{%
  \institution{Wuhan University}
  \country{China}
}

\author{Jin Song Dong}
\email{dcsdjs@nus.edu.sg}
\affiliation{%
  \institution{National University of Singapore}
  \country{Singapore}
}

\begin{document}

\title{Generating Project-Specific Test Cases with Requirement Validation Intention}

\begin{abstract}
Test cases are valuable assets for maintaining software quality.
State-of-the-art automated test generation techniques typically focus on maximizing program branch coverage or translating focal methods into test code.
However,
in contrast to branch coverage or code-to-test translation,
practical tests are written
out of the need to validate whether a requirement has been fulfilled.
Specifically, each test usually reflects a developer’s \textit{validation intention}
for a program function, regarding
(1) \textit{what is the test scenario of a program function?} and
(2) \textit{what is expected behavior under such a scenario?}
Without taking such intention into account,
generated tests are less likely to be adopted in practice.

In this work, we propose \tool,
which generates project-specific tests
given the description of validation intention.
The design is motivated by two insights:
(1) \textbf{rationale insight}: the description of validation intention 
regarding scenario description and behavioral expectation,
compared to coverage and focal code,
carries more crucial information about \ul{\textit{what to test}}; and
(2) \textbf{technical insight}: practical test code exhibits high duplication,
indicating that existing tests are highly reusable for \ul{\textit{how to test}}.
Therefore, \tool adopts a retrieval-and-edit manner.
First, given a focal code and a description of validation intention
consisting of a test objective with test precondition and expected results,
\tool retrieves a reusable test in the project as the test reference.
Then, \tool edits the test reference with an LLM regarding the validation intention toward the target test.
To guarantee that the target test can have a project-specific test prefix and a relevant test assertion,
\tool further explores the software project to identify \textit{crucial code facts} (i.e., relevant API/code to call and global variables to refer to in the test) as important context for the test generation.
We extensively evaluate \tool against four baselines (TELPA, DA, ChatTester, and EvoSuite) on 3,680 test cases from 12 open-source projects.
Compared to state-of-the-art baselines,
with a given validation intention,
\tool can
(1) generate tests far more semantically relevant to ground-truth tests
by
    (i) killing 28.1\% to 37.6\% more common mutants and
    (ii) sharing 16.9\% to 23.9\% more common coverage;
    and
(2) generate 23.7\% to 49.0\% more successful passing tests.
\end{abstract}



\received{20 February 2007}
\received[revised]{12 March 2009}
\received[accepted]{5 June 2009}

\maketitle

\input{1_introduction}

\input{2_motivating_example}

\input{3_empirical_study}

\input{4_approach}

\input{5_evaluation}

\input{6_discussion}

\input{7_related_work}

\input{8_conclusion}

\begin{acks}
This research is supported in part by the National Natural Science Foundation of China (62572300), the Minister of Education, Singapore (MOE-T2EP20124-0017, MOET32020-0004), the National Research Foundation, Singapore and the Cyber Security Agency under its National Cybersecurity R\&D Programme (NCRP25-P04-TAICeN), DSO National Laboratories under the AI Singapore Programme (AISG Award No: AISG2-GC-2023-008-1B), and Cyber Security Agency of Singapore under its National Cybersecurity R\&D Programme and CyberSG R\&D Cyber Research Programme Office, and partially by HUAWEI’s Al Hundred Schools Program using the Ascend AI technology stack. Any opinions, findings and conclusions or recommendations expressed in this material are those of the author(s) and do not reflect the views of National Research Foundation, Singapore, Cyber Security Agency of Singapore as well as CyberSG R\&D Programme Office, Singapore.
\end{acks}

\bibliographystyle{ACM-Reference-Format}
\bibliography{sample-base}

\end{document}

%% file: package.tex
\PassOptionsToPackage{table,dvipsnames}{xcolor}
\usepackage{colortbl}
\usepackage{amsmath,amsfonts}
\usepackage{algorithmic}
\usepackage{textcomp}
\usepackage{color}
\usepackage{booktabs}

\usepackage{enumitem}
\usepackage{xspace}
\usepackage{multirow}

\usepackage[linesnumbered,ruled,vlined]{algorithm2e}

\usepackage{graphicx}
\usepackage{listings}
\usepackage{xcolor}
\usepackage{enumitem}
\usepackage{amsmath}
\usepackage{float}
\usepackage{graphicx}
\usepackage{multirow}
\usepackage{caption}
\usepackage{subcaption}
\usepackage{xspace}
\usepackage{verbatim}
\usepackage{hyperref}
\usepackage{cleveref}
\usepackage{soul}
\usepackage{array}
\usepackage{adjustbox}
\usepackage{tabularx}
\usepackage{wrapfig}

\usepackage{diagbox}

\usepackage{tcolorbox}

\newcommand{\argmax}{\operatorname*{arg\,max}}

\usepackage{varwidth}
\newcommand{\coloredbox}[2]{%
    \colorbox{#1}{\begin{varwidth}[t]{\dimexpr\linewidth-2\fboxsep\relax}#2\end{varwidth}}%
}

\usepackage{tikz}
\newcommand{\stepcircle}[1]{%
    \tikz[baseline=-0.7ex]{
    \node[circle, fill=black, text=white, font=\sffamily\scriptsize\bfseries,
      inner sep=0.35pt, minimum size=0.75em] (char) {#1};
  }%
}

\definecolor{SoftRed}{RGB}{255, 228, 225}
\definecolor{SoftGreen}{RGB}{230, 255, 230}
\definecolor{SoftOrange}{RGB}{255, 229, 204}
\definecolor{SoftPurple}{RGB}{224, 224, 248}
\definecolor{SoftBlue}{RGB}{202, 240, 255}
\definecolor{SoftGray}{RGB}{237, 237, 237}
\definecolor{DeepGreen}{RGB}{0,100,0}
\definecolor{VioletCustom}{RGB}{138,43,226}

\newcommand\tool{IntentionTest\xspace}

%% file: 1_introduction.tex
\section{Introduction}
Software tests play a crucial role in ensuring the quality of both open-source and industrial software.
They can be used for requirement clarification \cite{mu2024clarifygpt, mu2023clarifygpt}, code review \cite{baum2016need, zhou2023generation}, and CI/CD \cite{campos2014continuous, gota2020continuous, fallahzadeh2023accelerating},
ensuring the reliability of software products.
The state-of-the-art software testing approaches can generally fall into the following categories:

\begin{itemize}[leftmargin=*]
  \item \textbf{Test Generation for Code Coverage.}
    Classical test generators~\cite{cadar2008klee,godefroid2005dart,pacheco2007randoop,lemieux2023codamosa,lin2021graph} such as
    Klee \cite{cadar2008klee}, Dart \cite{godefroid2005dart},
    EvoSuite \cite{fraser2011evosuite},
    Randoop \cite{pacheco2007randoop},
    and their variants \cite{lemieux2023codamosa, lin2021graph}
    consider test generation
    as a problem of maximizing branch or path coverage.
    To this end, the problem of test generation is transformed into a problem of constraint solving,
    with solutions such as static and dynamic symbolic execution \cite{cadar2008klee, braione2017combining, braione2018sushi, godefroid2005dart, sen2005cute} and search-based software testing \cite{fraser2011evosuite, arcuri2008search, braione2017combining, godoy2021enabledness, lin2021graph, lemieux2023codamosa, pacheco2007randoop, lin2020recovering}.
  \item \textbf{Test Generation as Code Translation.}
    With the emergence of language models,
    LLM-based test generators \cite{nie2023learning, tufano2020unit, kang2023large, dinella2022toga, chattester} consider the test generation as a special case of code generation.
    Given a focal method and an instruction prompt,
    those techniques use LLMs (e.g., ChatGPT) to inductively generate tests by translating the prompt, focal method, or both to the test code.
\end{itemize}

While existing LLM-based test generators have shown promising results,
practical test cases are not purely driven by coverage metrics or code-to-test translation.
Instead,
software developers write test cases by
validating the consistency between the requirement and its implementation.
Such a validation intention usually indicates
(1) the test scenario of the program function and
(2) the expected behavior of the program function.
For example, a program function to start a server (see \autoref{fig:focal-method}) can be tested under a scenario
where the option of using a thread pool.
Its behaviors are validated by checking whether the pool is created as expected.
Such intention can hardly be inferred from program branches (see the branchless code example in \autoref{fig:focal-method}).
However, it is crucial to guide how we write the target tests.

In this paper, we use the term validation intention to denote the explicit articulation of what requirement to validate and how its fulfillment should be observed.
Different from branch coverage, which focuses on structural exploration or code-to-test translation, 
which merely mirrors the focal method, validation intention captures the semantic link between requirements, usage scenarios, and expected behaviors.
It thus serves as the conceptual bridge from requirement specification to executable tests. 
Concretely, 
validation intention can be instantiated as a semi-structured description 
(objective, preconditions, and expected results),
but its essence lies in guiding test design 
towards meaningful requirement validation rather than superficial coverage.

According to our studies in a technical giant
with over 10,000 software developers,
regulated validation intention descriptions are adopted to write 67.8\% of system tests and 18.6\% of unit tests.
However, converting validation intention into test code is technically challenging
with the following knowledge gaps.

\begin{itemize}[leftmargin=*]
  \item \textbf{Implicit Project-Specific Test Idiom}:
    First, writing tests is more than just invoking the focal methods,
    which requires following specific project paradigms,
    including decisions such as
    ``\textit{How to construct the test inputs (e.g., constructor, factory, or singleton)?}'', ``\textit{What object to be mocked?}'', and
    ``\textit{What resource to release after the test?}''.
    As a result, additional project-specific code often appears before or after the invocation of the focal method in the test.

  \item \textbf{Implicit Project-Specific Constraints}:
    Second, test inputs can have very restrictive choices in the project.
    For example, in the Spark project \cite{spark}, the range of parameter with name \texttt{headerType} of \texttt{int} type can only be limited to a few choices (e.g., 0, 1, 2, and 3, each stands for a type of message header) instead of falling into the range of $-2^{n-1}$ to $2^{n-1}-1$ ($n$ is the bit number).
\end{itemize}

In this work, we propose \tool, which compiles the validation intention description into a \textit{project-specific} test.
The description of validation intention
consists of a test objective,
complemented with an optional test precondition and an expected result.
Based on the
empirical observation that test code is highly duplicated (see Section~\ref{sec:empirical-study}),
we design \tool to generate tests in a retrieval-and-edit manner.
Specifically, given (1) a focal method $m$ and (2) a validation intention description $desc$,
\tool generates tests in two stages, i.e.,
(1) in the \textbf{retrieval stage}, \tool constructs $m$ and $desc$ as a query to search for an existing test as a reference and
(2) in the \textbf{edit stage}, \tool reduces the test generation problem into a code-editing problem on the reference.
To edit the test reference into the correct target test,
we explore the entire project for the crucial code facts (e.g., crucial project-specific APIs)
based on
(1) their semantics to the validation intention and
(2) their historical relevance to the focal method (e.g., how often an API or a piece of code co-occurs with the invocation of the focal method).
The crucial facts can significantly mitigate the LLM hallucination for reliable edits on the test reference.
Finally, \tool iteratively refines the generated test, resolving compilation errors, execution errors, and assertion failures until the test passes or the maximum iteration is reached.

We extensively evaluate \tool against four baselines (i.e., TELPA~\cite{telpa}, DA~\cite{issta24_test_adaption}, ChatTester~\cite{chattester} and EvoSuite ~\cite{fraser2011evosuite}) on 3,680 tests from 12 open-source projects.
Compared to state-of-the-art approaches, i.e., TELPA, DA, and ChatTester,
with a given validation intention,
\tool can
(1) generate tests far more semantically relevant to the ground-truth tests by
(i) killing 28.1\% to 37.6\% more common mutants and
(ii) invoking up to 62.9\% more project-specific APIs;
(2) generate 36.7\% to 49.0\% more successful passing tests.
Moreover, \tool maintains its performance across both commercial LLMs (e.g., ChatGPT) and open-source LLMs (e.g., DeepSeek-V3.2).
Finally, we evaluate the performance of \tool across different granularities of the description of validation intention,
demonstrating \tool can still keep its performance
even when we restrict the description by only keeping the test objective within 50 words.

In summary, this work makes the following contributions:
\begin{itemize}[leftmargin=*]
  \item \textbf{Methodology.}
    To the best of our knowledge, \tool is the first solution to
    convert humans' validation intention (i.e., test requirement) to a project-specific test with both test prefix and oracle.
    \tool is an agentic test generator supporting multiple granularities of intention descriptions,
    with stable test generation performance.
  \item \textbf{Empirical Study.}
    We conduct an empirical study over 12 open-source projects, showing that test reuse is prevalent in well-established projects.
    This finding provides a strong empirical foundation for the design of retrieval-and-edit-based test generation solutions.
  \item \textbf{Dataset \& Tool.}
    We release the first benchmark where
    the focal code, test code, and validation intention (regarding the test objective, preconditions, and expected results) are available.
    We collect and curate 3,680 test cases across 12 projects to ensure high quality.
    Also, we deliver \tool as a prototype tool in the form of a VS Code extension,
    visualizing the stepwise process (see tool video in~\cite{tool_video}).
    The tool can facilitate practical test generation,
    particularly for well-established projects.
  \item \textbf{Evaluation.}
    We extensively evaluate \tool against state-of-the-art baselines on our benchmark.
    Results show that \tool can generate tests that are more semantically faithful to the validation intention while also outperforming baselines on general metrics, including mutation score, pass rate, and code coverage.
    Furthermore, we analyze the performance of validation intention across five levels of detail,
    demonstrating that it can still maintain the performance even when limited to a concise description of the test objective within 50 words.
\end{itemize}

Given the space limit, the demonstration video, tool source code, and more experimental results are available at \cite{dtester}.

%% file: 2_motivating_example.tex
\section{Motivating Example} \label{sec:example}

\autoref{fig:focal-method} shows a focal method in the Spark project \cite{spark},
which creates an embedded server, such as Jetty and Tomcat, in the Spark distributed system.
Generating its test cases with the state-of-the-art solutions (e.g., EvoSuite~\cite{fraser2011evosuite}, DA~\cite{issta24_test_adaption}, and ChatTester~\cite{chattester}) for such focal code as in \autoref{fig:target-test} is challenging for the following reasons:
\begin{itemize}[leftmargin=*]
  \item \textbf{A Branch Cannot Represent A Test Scenario/Intention.}
    Despite that the focal method \texttt{create()} is short and branchless (see \autoref{fig:focal-method}),
    there are many validation options to test its different scenarios.
    For example, the server can be created with different options of thread pool and configured cookies.
    Therefore, %
    full branch coverage can hardly indicate full scenario coverage.
    Without a specified validation intention (i.e., \textit{create a server with thread pool}) as guidance,
    it is challenging to define and generate a \textit{semantically} correct test.
  \item \textbf{Abundant Project Specifics.}
    As shown in \autoref{fig:target-test}, the ground-truth test contains project-specific test prefixes and oracles.
    First, the test requires specific mocking since
    initializing or instantiating certain objects can be time- and resource-consuming (Lines 3-6).
    Second, a specific mocking condition (Line 8) must be defined before calling the focal method \texttt{create()}.
    Third, the test scenario requires calling two APIs (Lines 12-13) after invoking the focal method to ignite the server with a specific configuration, ensuring that the created server is runnable.
    Finally, unlike common assertions that simply check for the existence of created objects, this test scenario verifies the invocation time and the number of interactions by the created server (Lines 15-16).
\end{itemize}

\lstset{
  language=Java,
  basicstyle=\ttfamily\fontsize{7pt}{7pt}\selectfont,
  keywordstyle=\color{blue},
  stringstyle=\color{VioletCustom},
  commentstyle=\color{blue},
  numbers=left,
  numberstyle=\tiny\color{black},
  stepnumber=1,
  numbersep=1pt,
  showspaces=false,
  showstringspaces=false,
  tabsize=4,
  breaklines=true,
  breakatwhitespace=false,
  showtabs=false,
  frame=single,
  linewidth=0.98\columnwidth,
  escapeinside={(*@}{@*)},
  captionpos=b,
  belowcaptionskip=-10pt,
  xleftmargin=13pt,
  framexleftmargin=5pt
}

\begin{lstlisting}[float=t, caption=Focal method needs to be equipped with test cases in Spark project \cite{spark}., label=fig:focal-method]
public EmbeddedServer create(Routes rm, StaticFilesConfiguration conf, ExceptionMapper em, boolean han) {
  MatcherFilter filter = new MatcherFilter(rm, conf, em, han);
  filter.init(null);

  JettyHandler handler = new JettyHandler(filter);
  handler.getSessionCookieConfig().setHttpOnly(httpOnly);
  return new EmbeddedJettyServer(serverFactory, handler).withThreadPool(threadPool);
} 
\end{lstlisting}

\begin{lstlisting}[float=t, caption={Target test corresponds to the focal method in Spark project. Codes in green are project-specific, which can hardly be inferred by LLMs learned from a general corpus.}, label=fig:target-test]
public void create_withThreadPool(){
    QueuedThreadPool pool = new QueuedThreadPool();
    JettyServerFactory jFactory = mock(...);
    StaticFilesConfiguration conf = mock(...);
    ExceptionMapper map = mock(...);
    Routes routes = mock(...);

    (*@\textcolor[rgb]{0.00,0.50,0.50}{\textbf{when(jFactory.create(pool)).thenReturn(new Server(pool));}}@*)

    EmbeddedJettyFactory eFactory = new EmbeddedJettyFactory(jFactory).withThreadPool((*@\textcolor[rgb]{0.00,0.50,0.50}{\textbf{pool}}@*));
    embeddedServer = eFactory.create(routes, conf, map, false);
    (*@\textcolor[rgb]{0.00,0.50,0.50}{\textbf{embeddedServer.trustForwardHeaders(true);}}@*)
    (*@\textcolor[rgb]{0.00,0.50,0.50}{\textbf{embeddedServer.ignite("localhost", 6758, null, 0, 0, 0);}}@*)

    (*@\textcolor[rgb]{0.00,0.50,0.50}{\textbf{verify(jFactory, times(1)).create(pool);}}@*)
    (*@\textcolor[rgb]{0.00,0.50,0.50}{\textbf{verifyNoMoreInteractions(jFactory);}}@*)
}
\end{lstlisting}

\begin{lstlisting}[float=t, caption={Test generated by ChatTester, missing project-specific pre- and post-setup.}, label=fig:chattester-example]
public void create_withThreadPool() {
    Routes routeMatcher = mock(Routes.class);
    StaticFilesConfiguration staticFilesConfiguration = mock(StaticFilesConfiguration.class);
    ExceptionMapper exceptionMapper = mock(ExceptionMapper.class);
    ThreadPool pool = new QueuedThreadPool();
    EmbeddedJettyFactory factory = new EmbeddedJettyFactory().withThreadPool(pool);

    EmbeddedServer server = factory.create(routeMatcher, staticFilesConfiguration, exceptionMapper, false);

    assertNotNull("EmbeddedServer should not be null", server);
}
\end{lstlisting}

\noindent\textbf{Coverage-based Test Generators}.
For such a branchless focal method,
traditional coverage-driven software testing lacks sufficient information to distinguish different test scenarios.
In addition, although the generated tests are runnable, they often omit crucial project-specific setup and, by design, cannot reflect the developer’s validation intention.
Given the space limit, interested audiences can check an example of
the EvoSuite-generated test at \cite{dtester}.

\noindent\textbf{LLM-based Test Generators (Code-to-Test Translation)}.
\autoref{fig:chattester-example} presents a test generated by
ChatTester~\cite{chattester}, a solution built on GPT-o1-mini.
Even when fed with detailed validation intention from \autoref{tab:test-desc} as its additional prompt,
ChatTester misses crucial project-specific pre- and post-conditions, including (1) specifying additional mocking behaviour (Line 8 in \autoref{fig:target-test}) and (2) starting the server with specific configurations and generating assertions with specific intents (Lines 12-16 in \autoref{fig:target-test}).

\input{tables/test_desc}

\begin{lstlisting}[float=t, caption={Referable test retrieved and edited by \tool towards the target test. The code in red is the test code to be modified, and the code in green is the expected test code.}, label=fig:reference-test]
(*@\textcolor{red}{\textbf{- public void create\_withoutHttpOnly()\{}}@*)
(*@\textcolor{DeepGreen}{\textbf{+ public void create\_withThreadPool()\{}}@*)
(*@\textcolor{DeepGreen}{\textbf{+}}@*)   (*@\textcolor{DeepGreen}{\textbf{QueuedThreadPool pool = new QueuedThreadPool();}}@*)
    JettyServerFactory jFactory = mock(...);
    StaticFilesConfiguration conf = mock(...);
    ExceptionMapper map = mock(...);
    Routes routes = mock(...);

(*@\textcolor{red}{\textbf{-}}@*)   Server server = new(*@\textcolor{red}{\textbf{ Server()}}@*);
(*@\textcolor{DeepGreen}{\textbf{+}}@*)   Server server = new (*@\textcolor{DeepGreen}{\textbf{Server(pool)}}@*);
    
(*@\textcolor{red}{\textbf{-}}@*)   when(jFactory.(*@\textcolor{red}{\textbf{create(100, 10, 10000)}}@*)).thenReturn(server);
(*@\textcolor{DeepGreen}{\textbf{+}}@*)   when(jFactory.(*@\textcolor{DeepGreen}{\textbf{create(pool)}}@*)).thenReturn(server);
    
(*@\textcolor{red}{\textbf{-}}@*)   EmbeddedJettyFactory eFactory = new EmbeddedJettyFactory(jFactory).(*@\textcolor{red}{\textbf{withHttpOnly(false})}@*);
(*@\textcolor{DeepGreen}{\textbf{+}}@*)   EmbeddedJettyFactory eFactory = new EmbeddedJettyFactory(jFactory).(*@\textcolor{DeepGreen}{\textbf{withThreadPool(pool)}}@*);
    
    embeddedServer = eFactory.create(routes, conf, map, false);
    embeddedServer.trustForwardHeaders(true);
    
(*@\textcolor{red}{\textbf{-}}@*)   embeddedServer.ignite("localhost", (*@\textcolor{red}{\textbf{6759, null, 100, 10, 10000}}@*));
(*@\textcolor{DeepGreen}{\textbf{+}}@*)   embeddedServer.ignite("localhost", (*@\textcolor{DeepGreen}{\textbf{6758, null, 0, 0, 0}}@*));

(*@\textcolor{red}{\textbf{-}}@*)   (*@\textcolor{red}{\textbf{assertFalse(server.getHandler().getSessionCookieConfig().isHttpOnly());}}@*)
(*@\textcolor{DeepGreen}{\textbf{+}}@*)   (*@\textcolor{DeepGreen}{\textbf{verify(jFactory, times(1)).create(pool);}}@*)
(*@\textcolor{DeepGreen}{\textbf{+}}@*)   (*@\textcolor{DeepGreen}{\textbf{verifyNoMoreInteractions(jFactory);}}@*)
\end{lstlisting}

To overcome the above challenges,
\tool adopts a retrieval-and-edit solution to parse a focal method and a description of validation intention (see \autoref{tab:test-desc}) to generate a test as shown in \autoref{fig:reference-test}.
Specifically, \tool
(1) retrieves referable tests relevant to the target test scenario and
(2) edits the referable test regarding the provided validation intention description.
The code in red is where the edit location shall happen,
and the code in green is the expected edit content in each location.
Note that, the performance of \tool is stable
if we only keep the section of test objective and expected results in \autoref{tab:test-desc},
which enables the programmers to spend minimal effort on writing the validation intention.

\noindent\textbf{Challenges of Knowledge Gap.}
To generate a test as shown in \autoref{fig:reference-test},
\tool needs to be aware of crucial code facts such as
the definition of \texttt{Jetty\-Server\-Factory.create(ThreadPool)}, \texttt{Server(Thread\-Pool)},
and \texttt{Embedded\-Jetty\-Factory.with\-Thread\-Pool(Thread\-Pool)} (see green text in \autoref{fig:reference-test})
from thousands of program elements in the project.
Without those facts,
LLM can easily hallucinate arbitrary non-existent APIs to invoke in the tests,
causing compilation errors.
However, the description of validation intention is in natural language form,
sometimes very different from the API names; 
thus, precisely mapping the description to concrete program elements is non-trivial.
In addition, it is also non-trivial to adopt the identified facts
into code edits on the test reference.

%% file: tables/test_desc.tex
\begin{table}[t]
\setlength{\abovecaptionskip}{0pt}
\setlength{\belowcaptionskip}{0pt}
\caption{
The validation intention description follows ISO/IEC/IEEE 29119 \cite{international_standard}.
\textit{Objective} section is mandatory, \textit{Preconditions} (optional) and \textit{Expected Results} sections are optional.
}
\vspace{4pt}
\label{tab:test-desc}
\centering
\footnotesize

\begin{tabularx}{\linewidth}{X}

    \toprule

    \# \underline{\textbf{Objective}}:\\ Tests creating an embedded server with a custom thread pool and checks if the server starts correctly. \\

    \midrule

    \# \underline{\textbf{Expected Results}}:\\ 1. Jetty server factory creates a server with the specified thread pool exactly once.\\
2. The server is initialized with the provided configurations and launched at port 6758 successfully.\\
3. No additional interactions with Jetty server factory occur beyond initial creation. \\
    \bottomrule
\end{tabularx}
\end{table}

%% file: 3_empirical_study.tex
\section{Empirical Study}\label{sec:empirical-study}

\begin{wrapfigure}[16]{r}{0.5\textwidth}
\centering
\vspace{-10pt}
\begin{minipage}{0.48\linewidth}
    \centering
    \includegraphics[width=\linewidth]{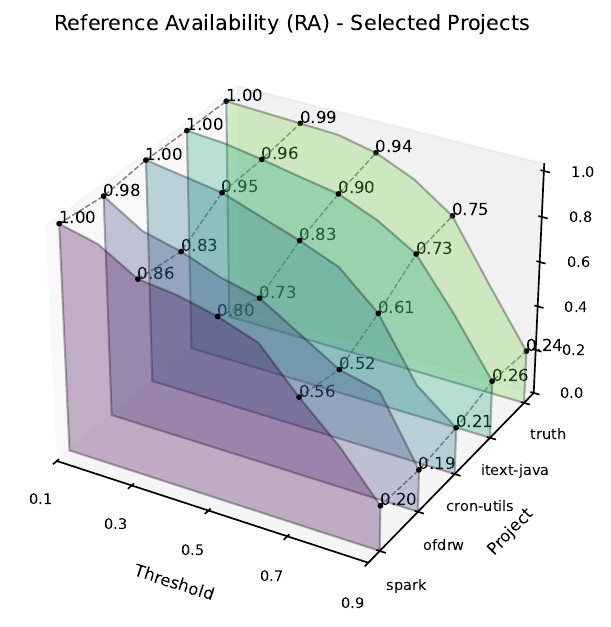}
    \small (a) Reference Availability
\end{minipage}
\hfill
\begin{minipage}{0.48\linewidth}
    \centering
    \includegraphics[width=\linewidth]{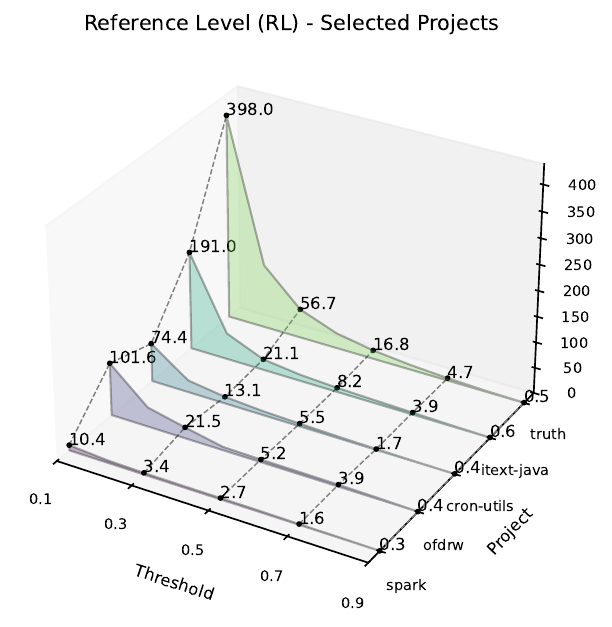}
    \small (b) Referability Level
\end{minipage}
\captionof{figure}{The results of the empirical test referability.
(a) and (b) present the RA and RL values and their trends across thresholds from 0.1 to 0.9 (as colored surfaces),
with an interval of 0.1.
Projects are sorted by the average number of candidates,
from the fewest (155.8 for \texttt{spark}) to the most (2191.9 for \texttt{truth}).
Even at high similarity threshold (e.g., 0.7), more than 50\% tests have reusable references.
}
\label{fig:empirical-study}
\vspace{-5pt}
\end{wrapfigure}

In this section, we report our empirical study on open-source projects to
evaluate our hypothesis that 
\ul{\textit{existing test cases are highly reusable assets for writing a new test case}}.
The validated hypothesis lays an important foundation for our retrieval-and-edit design of \tool.

\subsection{Setup}
\label{sec:empirical_setup}

\noindent\textbf{Dataset.}
We collect 12 diverse popular open-source projects,
spanning over web development, image processing, mobile development, and document generation conversion,
each with over 100 GitHub stars and forks.
For the initial dataset, $Dataset\text{-}FIX$, we snapshot a recent commit for each project to extract pairs of tests and their corresponding focal methods.
This results in a total of 3,680 tests, with individual projects having between 51 and 1,099 tests.
Each test has on average 25 lines of code (LoC), ranging from 9 to 124 LoC.
Each focal method has on average 12 LoC, ranging from 3 to 331 LoC.

To support the empirical study's temporal requirements, we further derived $Dataset\text{-}TEM$. 
For each test $t$ in $Dataset\text{-}FIX$, we reverted the project to the specific version where $t$ was first created. 
We then identified all other tests existing at that point in time as potential retrieval candidates for $t$. 
Across the 12 projects, the average number of candidates per test ranges from 74.0 to 2191.9.
Detailed statistics for both $Dataset\text{-}FIX$ and $Dataset\text{-}TEM$ are available in \cite{dtester}.

\noindent\textbf{Metrics.}
We design two metrics:
(1) \textbf{reference availability} for how likely a test can find its reference in its project? and
(2)  \textbf{reference level} for how many referable tests can a test have?
Specifically, given a similarity threshold $th$ and test similarity function $sim(. ,.)$, we evaluate test referability as follows:
\begin{itemize}[leftmargin=*]
  \item \textbf{Reference availability}:
    Given the set of tests, $T$, we compute reference availability $RA_{th}(T)$ as:
    \begin{equation}\label{eq:ra}
      RA_{th}(T) = \frac{\sum_{t\in T} \textbf{1}(sim(t, T\setminus \{t\}) > th)}{|T|},
    \end{equation}
    where $sim(t, T\setminus\{t\}) = \argmax_{t^*\in T\setminus\{t\}} sim(t, t^*)$.
    Intuitively, $RA_{th}(T)$ measures the ratio of tests in $T$ that have at least one non-identical test whose similarity exceeds $th$.
  \item \textbf{Referability level}:
    Given the set of tests, $T$, we calculate referability level $RL_{th}(T)$ as:
    \begin{equation}\label{eq:rl}
      RL_{th}(T) = \frac{\sum_{t\in T} count(sim(t, T\setminus \{t\}) > th)}{|T|}.
    \end{equation}
\end{itemize}

$RL_{th}(T)$ evaluates the average number of referable tests per test.
We use BM25 with normalization as the similarity metric,
and vary the threshold from 0.1 to 0.9 (in increments of 0.1) to assess how different similarity cutoffs affect the results.

\subsection{Results}

\autoref{fig:empirical-study} shows the results of the empirical study on test referability.
Overall, all 12 projects exhibit very high reference availability (RA) and referability level (RL).
In \autoref{fig:empirical-study}, we select 5 from the 12 projects based on the average number of candidates per test: \texttt{spark}~\cite{spark},
\texttt{ofdrw}~\cite{ofdrw},
\texttt{cron-utils}~\cite{cron-utils},
\texttt{itext-java}~\cite{itext},
and \texttt{truth}~\cite{truth},
with 155.8, 360.2, 475.8, 1279.4, and 2191.9 candidates, respectively.
In \autoref{fig:empirical-study}(a) and \autoref{fig:empirical-study}(b),
we show how the RA and RL value of each project under different thresholds,
the higher the area under the curve, the more RA and RL value.
For example in \autoref{fig:empirical-study}(a),
the \texttt{itext-java} project,
73\% of the tests have references at the similarity threshold of 0.7.
In \autoref{fig:empirical-study}(b),
even for the small project \texttt{spark}, $RL_{0.7}$ reaches 1.6,
which means a target test has close to two references.
Moreover, similar to the trend observed in $RA$, $RL$ also exhibits an increase as the number of historical candidates grows.
More details can be found in \cite{dtester}.

As a result, we conclude that
(1) test referability in an open-source project is prevalent and
(2) the more established a project, the more likely its tests can be referred to each other.

%% file: 4_approach.tex
\section{Approach}\label{sec:approach}

\begin{figure*}[t]
  \centering
  \includegraphics[width=0.85\textwidth]{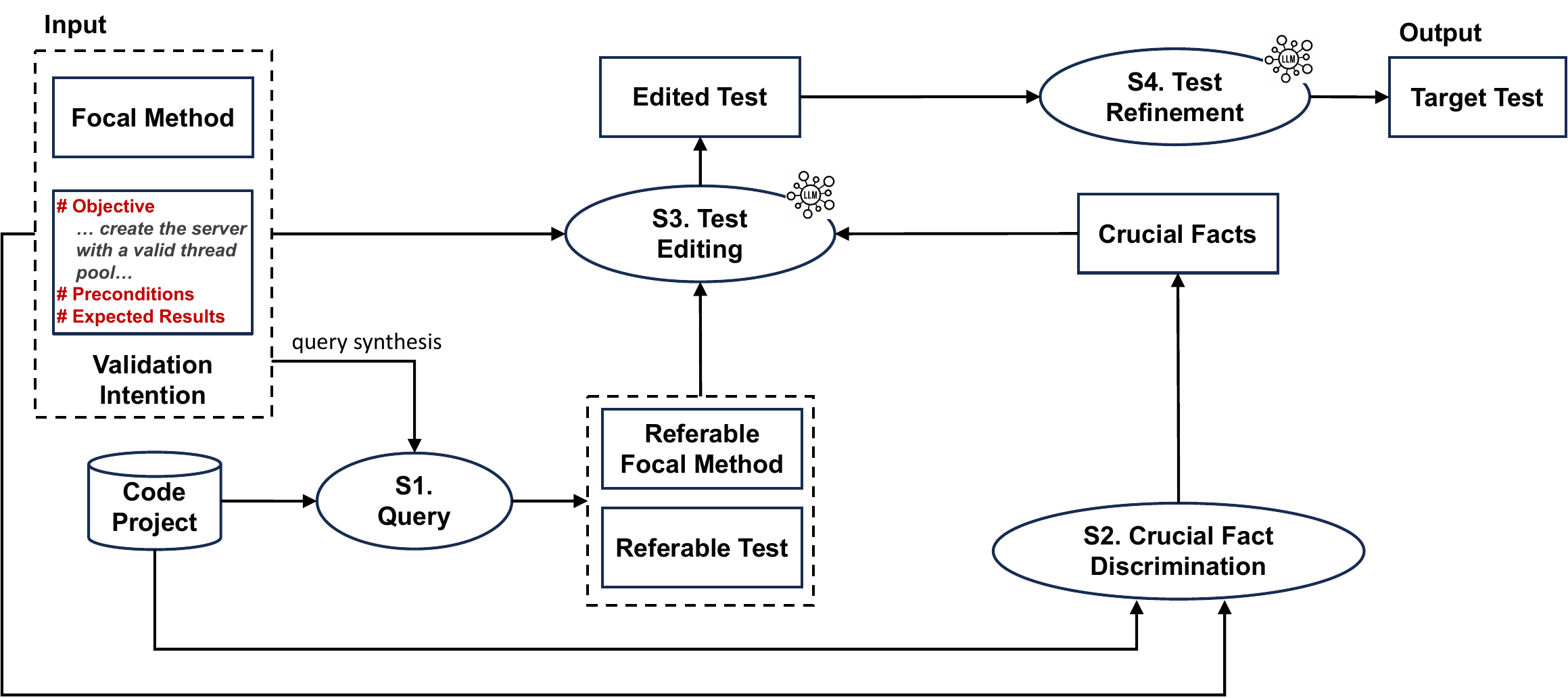}
  \caption{Overview of \tool: Given a focal method and a validation intention description of its test, \tool derives the test in a retrieval-and-edit manner.
  }\label{fig:overview}
  \vspace{-12pt}
\end{figure*}

\noindent \textbf{Problem Statement.}
Given a software project containing a set of tests $T$ and a set of functions $M$,
we assume each test $test\in T$ has a corresponding focal method $m\in M$ to test.
The validation description $desc_{tar}$ is defined as a tuple $(obj, precondition, expectation)$ where
(1) $obj$, $precondition$, and $expectation$ are textual description,
(2) $obj$ describes the test scenarios, $precondition$ describes what needs to be satisfied before running the focal method, and $expectation$ describes verifiable program behaviors, and
(3) $precondition$ can be $\epsilon$ (i.e., empty string).
Given a target focal method $m_{tar}$ to be tested, and its validation description $desc_{tar}$,
we generate the target test for $m_{tar}$, denoted as $test_{tar}$, regarding $desc_{tar}$.
Note that $desc_{tar}$ is a semi-structured description consisting of a test objective,
optionally complementary with test precondition and expected results.

\noindent \textbf{Approach Overview.}
\autoref{fig:overview} shows our approach consisting of the following stages:
\begin{itemize}[leftmargin=*]
  \item \textbf{Stage 1 (Referable Test Retrieval):}
    \tool converts $m_{tar}$ and $desc_{tar}$ into a query to retrieve a ranked list of method-test pairs,
    each consisting of a focal method $m_{ref}$ and its test $test_{ref}$, from the project.
    $test_{ref}$ is also called as a referable test to edit into the target test $test_{tar}$.
  \item \textbf{Stage 2 (Crucial Facts Discrimination):}
    Based on the validation intention $desc_{tar}$, \tool explores the project, following a variety of program relationships (e.g., caller, callee, declaration, etc.),
    for identifying crucial facts, $facts$, to edit the referable test.
  \item \textbf{Stage 3 (LLM-based Test Editing):}
    \tool constructs a prompt, incorporating general information such as $desc_{tar}$ and $m_{tar}$, as well as the project-specific information like $m_{ref}$, $test_{ref}$, and $facts$.
    Then, \tool prompts the LLM to edit $test_{ref}$ towards $test_{tar}$.
  \item \textbf{Stage 4 (LLM-based Self Refinement):}
    Once $test_{tar}$ is derived, \tool validates whether $test_{tar}$ compiles and executes successfully against $m_{tar}$.
    If errors occur, \tool iteratively refines $test_{tar}$ to correct compilation and execution errors based on error messages and $facts$.
\end{itemize}

\tool iterates these stages until $test_{tar}$ successfully passes, or a maximum number of iterations is reached.

\subsection{Referable Test Retrieval}\label{sec:retrieval}
Given $m_{tar}$ and $desc_{tar}$,
\tool retrieves from the project
a referable focal method $m_{ref}$ and
its referable test $test_{ref}$ to edit.
Intuitively, the fewer editing efforts to transform $test_{ref}$ into the target test $test_{tar}$,
the higher the likelihood of accomplishing the editing.
Technically, given a code similarity metric, denoted as $sim(. , .)$, we ideally have $test_{ref} = \argmax_{test \in T}$ $sim(test, test_{tar})$.
However, $sim(test, test_{tar})$ is not computable when we retrieve $test_{ref}$, as $test_{tar}$ is unknown.
Therefore,
we define a referability estimation function ($REF$) on the focal method $m_{tar}$, validation intention $desc_{tar}$, candidate referable method $m$, and candidate referable test $test$
so that the similarity between $test$ and $test_{tar}$ can be estimated. %
In this work, given a pair of $m$ and $test$,
we heuristically estimate the test referability by
(1) the similarity between $m_{tar}$ and $m$;
(2) the similarity between $desc_{tar}$ and the validation intention of $test$, denoted as $test.desc$:
{
\small
\begin{align}\label{eq:ref}
  REF(m_{tar}, desc_{tar}, m, test) &= \alpha \cdot sim(m_{tar}, m) + (1-\alpha) \cdot sim(desc_{tar}, test.desc).
\end{align}
}

In this equation, we let $\alpha \in (0, 1)$.
As for the similarity between $m_{tar}$ and $p.m$,
we use a normalized BM25 score (with min-max normalization).
Regarding the similarity between $desc_{tar}$ and $test.desc$,
we calculate the cosine similarity between their embeddings that are obtained using a lightweight embedding model, i.e., a pretrained CodeT5+ Embedding model~\cite{codet5}.
Note that, the summarized test objective, test precondition, and expected results allow us to compare tests in a more precise manner.
Finally, the pair $(m, test)$ with the highest REF value is selected as the reference.

\subsection{Crucial Facts Discrimination}
\label{sec:crucial-facts}

Given the referable test $test_{ref}$ and the validation intention $desc_{tar}$,
we then design an agentic project-wise editing approach to edit $test_{ref}$ to the target test.
Specifically, the LLM agent needs to be aware of
the program entities in the project for editing the test (see the code in green in \autoref{fig:reference-test}).
For example, in \autoref{fig:reference-test}, \tool shall identify
(1) \texttt{QueuedThreadPool} is the only appropriate implementation of the \texttt{ThreadPool} interface (see Line 3 in \autoref{fig:reference-test});
(2) the overloaded method \texttt{Server(ThreadPool)} should be used to replace \texttt{Server()}; and
(3) the overloaded method \texttt{JettyServerFactory.create(ThreadPool)} should be used to replace \texttt{JettyServerFactory.create(int, int, int)}.
We call such program entities (along with their code) as \textit{crucial fact} for the LLM agent to edit the referable test.

\begin{figure}
    \centering
    \includegraphics[width=\linewidth]{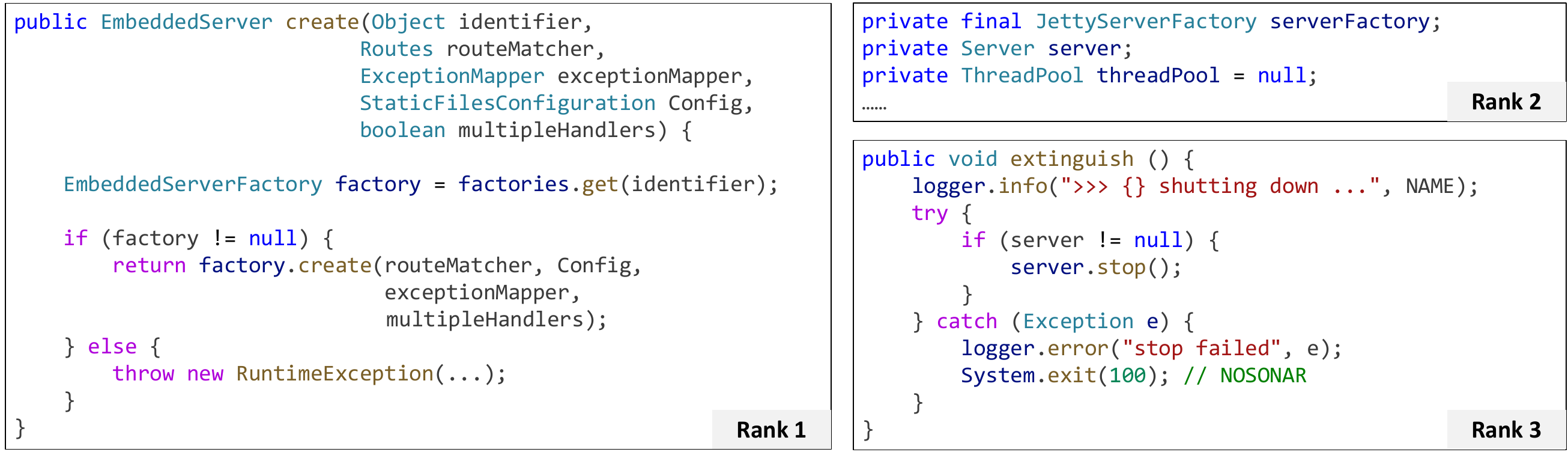}
    \caption{Taking the description of \autoref{tab:test-desc} as query (i.e., \textit{Tests creating an embedded server...}), all the top-3 retrieved program elements by embedding similarity are irrelevant.
    See a relevant program element in \autoref{fig:crucial-fact}.
    }
    \label{fig:fact_rank}
\vspace{-8pt}  
\end{figure}

\noindent\textbf{Challenge.}
However, a project can contain thousands of program entities while only a very small subset is relevant.
\autoref{fig:fact_rank} shows a naive embedding-based approach to retrieve
program elements regarding the intention description,
As a result, none of the top-3 elements is relevant,
which indicates that textual similarity is not sufficient for retrieving
the most relevant program entities.
False positive crucial facts can mislead our LLM agent to generate incorrect test code.

To address this, we first build a code graph where a node indicates a program entity (e.g., \textit{class}, \textit{method}, and \textit{field}), and an edge indicates their relationship (e.g., \textit{call} and \textit{define}),
for exploring crucial facts by discriminating program entities regarding both their \textit{semantic} and \textit{historical relevance}
to the focal method $m_{tar}$ and the validation intention $desc_{tar}$.

\subsubsection{Code Graph Exploration}
\label{sec:graph-explore}
The goal of code graph exploration is to collect facts relevant to
the focal method $m_{tar}$,
the intention description $desc_{tar}$, and
the referable test $test_{ref}$ for the target test $test_{tar}$ to adapt with.
Given $m_{tar}$, $desc_{tar}$, and $test_{ref}$,
\tool starts its exploration from both $m_{tar}$ and $test_{ref}$, than traversing the project's code graph.
In the graph,
each node represents a program entity, including \textit{class}, \textit{interface}, \textit{method}, and \textit{field};
each edge represents a relationship between entities, including \textit{define}, \textit{call}, \textit{param}, \textit{overload}, \textit{implement}, and \textit{extend}.
Note that, for each directed edge (e.g., call), we also derive a new edge as its reverse-direction relation (e.g., called by).
The maximum exploration depth from $m_{tar}$ and $test_{ref}$ is bounded by a user-defined threshold.

\subsubsection{Fact Discrimination}
\label{sec:disciminate}

For each visited program entity,
we evaluate its relevance by calculating its semantic and historical relevance to the focal method $m_{tar}$ and
the validation intention $desc_{tar}$.
While semantic relevance is still calculated based on embedding similarity,
we calculate historical relevance based on
how likely a program entity can be used together with the focal method.
Intuitively,
we consider a test case of a focal method $m_{tar}$ as a special case of code usage examples.
Therefore, to find the crucial facts for writing the test of $m_{tar}$,
we check all the existing usage examples in the project
for the frequent co-occurring entities of $m_{tar}$.
Technically, given a set of candidate facts $F$ (by traversing the code graph) and focal method usages $U$,
we rank $F$ by the likelihood of each candidate $f\in F$ by aggregating their quantified semantic and historical relevance.
Then, the top $k$ candidates are selected as the crucial facts set $F_{crucial}$.

\paragraph{Semantic Relevance Measurement.}
Given the intention description $desc_{tar}$ (e.g., \textit{``...creating an embedded server with a custom thread pool...''} in \autoref{tab:test-desc}) and a program entity $c$ ($c$ includes both code signature and body),
we use a pre-trained code embedding model (e.g., CodeT5+ \cite{codet5})
to have their embeddings, denoted as $e(desc_{tar})$ and $e(c)$.
Then we take their cosine similarity, i.e., $cos(e(desc_{tar}), e(c))$, as their semantic relevance.

\begin{figure}
    \centering
    \includegraphics[width=\linewidth]{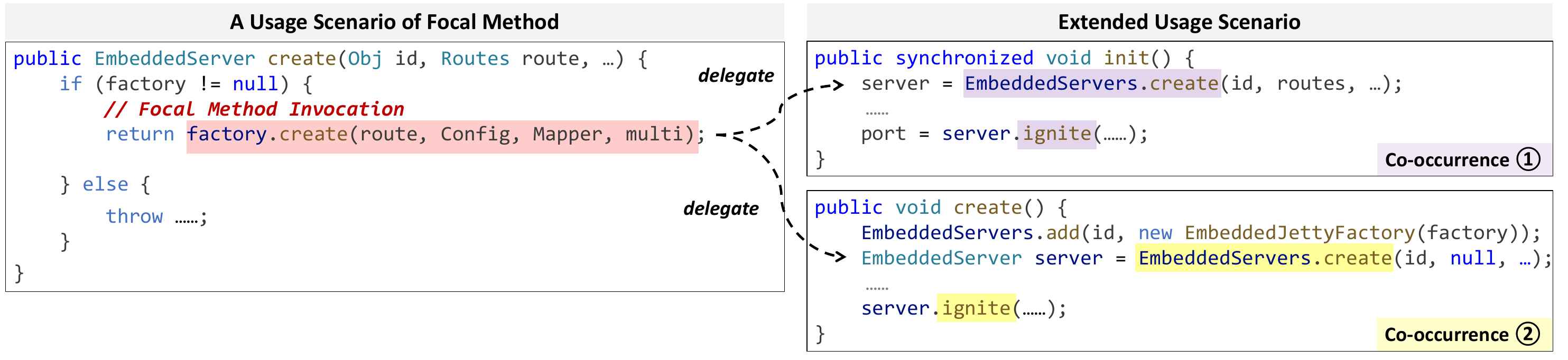}
    \caption{An example of extending usage scenarios.
    By discovering the method delegation relation,
    we can extend more usage scenarios to find the co-occurring program entities (a.k.a., \texttt{ignite()}) with the focal method (a.k.a, \texttt{create(...)}).}
    \label{fig:historical}
    \vspace{-5pt}
\end{figure}

\begin{figure}
    \centering
    \includegraphics[width=\linewidth]{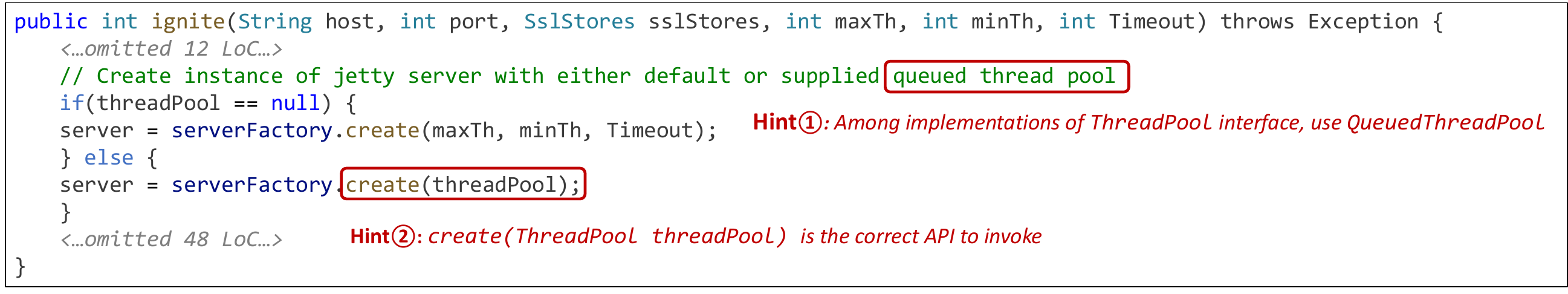}
    \caption{Crucial fact \texttt{ignite()} provides two hints for editing the test case shown in \autoref{fig:reference-test}.}
    \label{fig:crucial-fact}
    \vspace{-5pt}
\end{figure}

\paragraph{Historical Relevance Measurement.}
To mitigate the false positives and false negatives by embedding similarity,
we measure the historical relevance of a program entity to the focal method,
regarding their co-occurring likelihood.
To this end, we take two steps, i.e., co-occurrence identification and weighted co-occurrence calculation.

\noindent\textbf{Co-occurrence Identification.}
We first identify all the usage scenarios of the focal methods $m_{tar}$,
where each scenario is a method invoking $m_{tar}$.
Given a usage scenario $s_i$ of $m_{tar}$, we call the set of all its invoking program elements
as a \textbf{co-occurring set} to $m_{tar}$,
denoted as $co\_occur(s_i, m_{tar}) = \{e_1, e_2, ..., e_k\}$
where $e_i$ is a \textbf{co-occurring element} with $m_{tar}$.

In addition, we further expand the usage scenarios regarding the method delegation relation as shown in \autoref{fig:historical},
where the left rectangle represents a usage scenario of the focal method \texttt{create()} (in red background).
In this example,
the usage scenario of \texttt{create(Obj, Routes, ...)} can be
extended to the usage scenario of \texttt{init()} and \texttt{create()} (see the right rectangles in \autoref{fig:historical})
because the focal method \texttt{create()} is a \textit{delegation} of its caller method \texttt{create(Obj, Routes, ...)}.
In other words, the method \texttt{create()} and the method \texttt{create(Obj, Routes, ...)} 
have strong functional coherence despite their syntactic difference.
Here, we define a method $m_i$ is the \textbf{delegation} of a method $m_j$ if (1) $m_j$ calls $m_i$ and (2) $m_j$ takes the returned variable of $m_i$ as the returned variable.
Then, for each delegated method $m_d$ of the focal method $m_{tar}$,
we can take the co-occurring set of $m_d$
as the expanded co-occurring set of $m_{tar}$.

\noindent\textit{Example.}
For the focal method \texttt{create(...)} in \autoref{fig:historical},
we can have its original co-occurring set $cset_1 = \emptyset$ as there is no co-occurring program entities to \texttt{create(...)}.
Nevertheless, it has two expanded co-occurring sets,
from \texttt{init()} and \texttt{create()} respectively,
i.e., $cset_2 = \{$\texttt{ignite(...)}$\}$ and
$cset_2 = \{$\texttt{add}, \texttt{ignite(...)}$\}$.

\noindent\textbf{Weighted Co-occurrence Calculation.}
For each co-occurring element $e_i$, we calculate its co-occurring probability to $m_{tar}$ by:
\begin{align}
    occu_i=\sum_{cset_j \in CSet}\frac{sim(cset_j, desc_{tar})\cdot \textbf{1}(e_i\in cset_j)}{|CSet|},
\end{align}
where
(1) $\textbf{1}(e_i\in cset_j)$ returns 1 if the co-occurring element $e_i$ is in a co-occurring set $cset_j$, and 0 otherwise,
(2) $U$ is the set of co-occurring sets, and
(3) the embedding similarity (i.e., by cosine similarity) between a co-occurring set and the validation intention is used to weight each co-occurrence.
Finally, the likelihood of $f_i$ being a crucial fact is calculated as:
\begin{align}
    \mathcal{L}_i = \beta \times sim_i + (1-\beta) \times occu_i,
\end{align}
where $0<\beta<1$.
Finally, \tool selects program entities with their likelihood above a user-defined threshold
as the crucial fact.

\input{generation-prompt}

\subsection{Test Editing and Refinement}
\label{sec:test-adaption}

\autoref{tab:prompt-template} presents the prompt template for the LLM to edit the referable test,
which includes $m_{tar}$, $desc_{tar}$, $test_{ref}$, and $F_{crucial}$.
In the prompt, we parameterize some instructions such as \textit{Target Focal Method} and \textit{Referable Test Case}.
In addition, we enforce the LLM to focus on collected crucial facts (i.e., \textit{Crucial Project Knowledge}).

Given an edited test,
\tool compiles and executes the test, collecting feedback messages from the compiler and interpreter.
\tool extracts error messages and filters out irrelevant messages that involve files outside the project (e.g., code files from dependency packages).
Then, \tool constructs a prompt to refine the edited test, addressing the errors.
The prompt for refinement is based on the prompt for test editing with modifications: (1) adding the edited test and extracted error messages, and (2) changing ``\#Instruction'' part to request revision of the edited test.
\tool iteratively refines the edited test until it passes successfully or a maximum number of iterations is reached (e.g., 4 rounds).
More details about the prompts are available at \cite{dtester}.

%% file: generation-prompt.tex
\begin{table}[t]
\setlength{\abovecaptionskip}{0pt}
\setlength{\belowcaptionskip}{0pt}
\caption{
The prompt template for test editing. The content in red text is to be filled in interactively.
}
\vspace{4pt}
\label{tab:prompt-template}
\centering
\footnotesize

\begin{tabularx}{\linewidth}{X}

    \toprule

    \#\underline{Target Focal Method}: \coloredbox{SoftRed}{[code of target focal method]} \\

    \#\underline{Target Focal Method Context}: \coloredbox{SoftRed}{[skeleton of target focal file]} \\

    \#\underline{Target Test Case}: // A JUnit \coloredbox{SoftRed}{[version]} test case to be generated  \\

    \#\underline{Target Validation Intention Desc}:  \coloredbox{SoftRed}{[Objective, Preconditions, and Expected Results]}  \\

    \#\underline{Referable Test Case}: \coloredbox{SoftRed}{[code of referable test]} \\

    \#\underline{Crucial Project Knowledge}: \coloredbox{SoftRed}{[list of crucial facts]} \\

    \midrule

    \#\underline{Instruction}: Please generate ONE \#Target Test Case\# for \#Target Focal Method\# by strictly following \#Target Validation Intention Desc\# and referring to \#Referable Test Case\# and \#Relevant Project Information\#....\\

    \#\underline{Requirements}: Your final output must contain only ONE test method annotated `@Test' and strictly adhere to the following format: 
    (1) Begin with the exact prefix: ``\texttt{\`{}\`{}\`{}}package ''. 
    (2) End with the exact suffix: ``\texttt{\`{}\`{}\`{}}''. \\
    \bottomrule
\end{tabularx}

\end{table}

%% file: 5_evaluation.tex
\section{Evaluation} \label{sec:evaluation}

We evaluate \tool with three research questions:

\noindent\textbf{RQ1 (Overall Performance):} How effective is \tool in generating project-specific tests compared to the state-of-the-art test generators?

\noindent\textbf{RQ2 (Granularity of Validation Intention):} How does the granularity of intention description affect \tool's performance?
Or, how many details/efforts do the programmers need to provide to generate acceptable tests?

\noindent\textbf{RQ3 (Ablation Study):} How do referable tests (i.e., retrieval) and crucial facts (i.e., crucial fact discrimination) contribute to overall performance?

\input{rq1.tex}

\input{rq2.tex}

\input{rq3.tex}

%% file: rq1.tex
\subsection{Overall Performance (RQ1)}\label{sec:overall-performance}

\subsubsection{Dataset}\label{sec:data-leakage}
We use the 3,680 test cases, as the ground-truth, from 12 open-source repositories (see Section \ref{sec:empirical-study}) 
to evaluate the performance of \tool.

\noindent \textbf{Validation Intention Generation.}
Due to the scarcity of explicit requirement documentation in open-source repositories, we synthesized validation intentions to facilitate a robust evaluation. 
To ensure experimental diversity and rigor, we generated these descriptions via two strategies:
\begin{itemize}[leftmargin=*]
    \item \textbf{LLM-Inferred}:
    For each focal method, we prompt GPT-o1-mini to propose candidate validation intentions without access to the corresponding test. 
    We then align each intention to an existing test using an LLM-based matching step; we discard cases where no clear correspondence is found. This yields 2,536 test–intention pairs.
    \item \textbf{Human-Written}: We recruited six experienced developers to author validation intentions for 40 focal methods of 9 projects based only on the code implementation.
    The participants then mapped these intentions to relevant tests, resulting in a curated subset of 86 tests with human-validated intention descriptions.
\end{itemize}

\noindent \textbf{Commit Timeline Reconstruction.}
To prevent temporal leakage in retrieval, we reconstruct each test's creation time and restrict retrieval candidates to tests whose last modification predates the target test's creation.

\noindent \textbf{Mitigation of LLM's Memorization Effect.}
To avoid the inflated performance caused by the fact that LLMs could sometimes already \textit{memorize} the target test code,
we adopt an in-context unlearning technique \cite{takashiro2024answer}
for LLMs to
``forget'' the test code details.
Specifically, once we find that a naive prompt (e.g., just given the project name) can generate textually similar test code to that of a target test,
we follow Pawelczyk et al's in-context unlearning approach \cite{pawelczyk2023context} by introducing an unlearning prompt forcing LLM to ignore the whole knowledge of the project.

\subsubsection{Metrics}
\label{sec:metrics}
We evaluate \tool against the baselines upon the following metrics for
(1) the syntactic quality of tests and
(2) the semantic relevance of tests to the validation intention.
\begin{itemize}[leftmargin=*]
  \item \textbf{Compilation Failure, Execution Failure, Assertion Failure, and Successful Pass}:
  We track the rates of compilation failures, execution failures, assertion failures, and successful passes of the generated tests.
  An execution failure refers to test crashes at runtime, e.g., loading a nonexistent resource.
  An assertion failure refers to test's failing its assertion.
  \item \textbf{Common Mutation Score}:
  To quantify how much a generated test satisfies the given validation intention,
  we introduce the \textit{common mutation score} (CMS) to compare the generated test and the ground-truth test to recover.
  Given a generated test $t$ which can pass the assertion, we compare $t$ with the corresponding ground-truth test $t^*$, let $S_t$ and $S_{t^*}$ denote the sets of mutants each test kills.
  CMS is calculated via the Jaccard Index over these sets: $CMS(t)=JI(S_t, S_{t^*})=\frac{|S_t\cap S_{t^*}|}{|S_t\cup S_{t^*}|}$,
  where $CMS(t)$ ranges from 0 (totally different mutant-killing behaviour) to 1 (identical mutant-killing behaviour).
  Higher CMS values indicate stronger \textit{intention alignment} between $t$ and $t^*$.
  The performance of a test generator in CMS is measured by the average CMS over all passing tests: $CMS(T)=\frac{1}{|T|}\sum_{t\in T}CMS(t)$.
  Note that, different approaches can have different performance in generating the assertion-passing tests.
  Therefore, we further introduce $CMS_{pair}$ to compare \tool with a baseline on their shared assertion-passing tests.

  \item \textbf{Line Coverage Overlap}:
  For each generated test that executes successfully, we compute line coverage on the focal method and compare it with the ground-truth test: (1)\textit{Exact match} indicates identical covered-line sets; (2) \textit{Common coverage ratio} is the Jaccard similarity between covered-line sets.
  Similarly to CMS, the proportion of generated tests that achieve these is used to measure alignment with validation intention.
  \item \textbf{CodeBLEU}: We also use CodeBLEU~\cite{ren2020codebleu} to evaluate the similarity between the generated and ground-truth test cases.
\end{itemize}

\subsubsection{Baselines}

We choose \stepcircle{1} ChatTester~\cite{chattester} and \stepcircle{2} TELPA~\cite{telpa} as state-of-the-art agentic baselines
for its outstanding performance among LLM-based test generators.
To ensure a fair comparison, we upgraded the LLM used by them to o1-mini.
TELPA is designed to generate tests for all callable focal methods in a project, and a focal method can have multiple generated tests.
For a fair comparison, we select the best-performing generated test that has the highest CMS value among the generated tests for the focal method.
We choose \stepcircle{3} DA~\cite{issta24_test_adaption} as a state-of-the-art training-based baseline.
Following DA's settings, we fine-tune its CodeT5 model on our dataset and evaluate its performance on our benchmark.
We split the tests in each repository into two sets,
each used as the training and the testing set for once.
This ensures that all tests can be evaluated.
Moreover, we choose \stepcircle{4} EvoSuite~\cite{fraser2011evosuite} as the state-of-the-art search-based baseline,
since its authors are actively incorporating new search strategies in the tool.
As EvoSuite generates multiple tests for a focal method, we select a single test whose code coverage is most similar (measured by F1 score) to the ground-truth test. We do not use CMS for selection due to the trivial assertions of EvoSuite's generated tests.

We do not include IntUT~\cite{IntUT} as a baseline, despite it having a seemingly similar idea to \tool.
IntUT uses generated validation intentions to guide LLMs in test generation, where \textit{each intention targets a specific branch} in the focal method by specifying input parameters and expected outputs.
As a result, IntUT is primarily designed to maximize code coverage.
Moreover, its expected outputs are derived directly from the focal method, which assumes that the focal method returns constant values.
Finally, neither the source code nor the dataset used by IntUT is publicly available, making a fair and reproducible comparison infeasible.

\subsubsection{Results}
\input{tables/rq1-summary}

\autoref{tab:rq1-summary} summarizes all metrics with average performance across projects in the two datasets (i.e., a dataset using LLM-Inferred validation intention and another one using Human-Written validation intention).
The column ``Ours (DS)'' reports results run upon DeepSeek-V3.2, and ``Ours (O1)'' upon o1-mini.
In the experiment, \tool can find a referable test in 69.01\% and 75.32\% of cases of test generation on ``LLM-Inferred'' and ``Human-Written'' datasets, respectively.
A detailed breakdown of performance comparison across key metrics is available at \cite{dtester}.
Overall, \tool outperforms the LLM-based baselines and performs on par with EvoSuite in general metrics (i.e., compilation failure, execution failure, assertion failure, and successful pass).

\noindent\textbf{Performance on General Metrics}
We observe that DA, trained upon a small language model (i.e., CodeT5),
struggles with syntax errors~\cite{issta24_test_adaption}.
Moreover, EvoSuite leads the performance of generating passing test cases on most projects as it adopts regression oracles.
\tool significantly outperforms TELPA, ChatTester, and DA in success pass rate by 27.88\%, 23.66\%, and 49.00\% on ``LLM-Inferred'' dataset. 
The results on ``Human-Written'' dataset consistently demonstrate the superiority of \tool, with 37.48\%, 43.48\%, and 50.22\% over TELPA, ChatTester, and DA, respectively.
Upon investigation, we find that compilation and execution errors by the baselines (i.e., TELPA, DA, and ChatTester) are often caused by missing project-specific details.
For example, many compilation errors stem from 
calling non-existent methods or
missing project-specific packages.

\begin{figure}
    \centering
    \includegraphics[width=\linewidth]{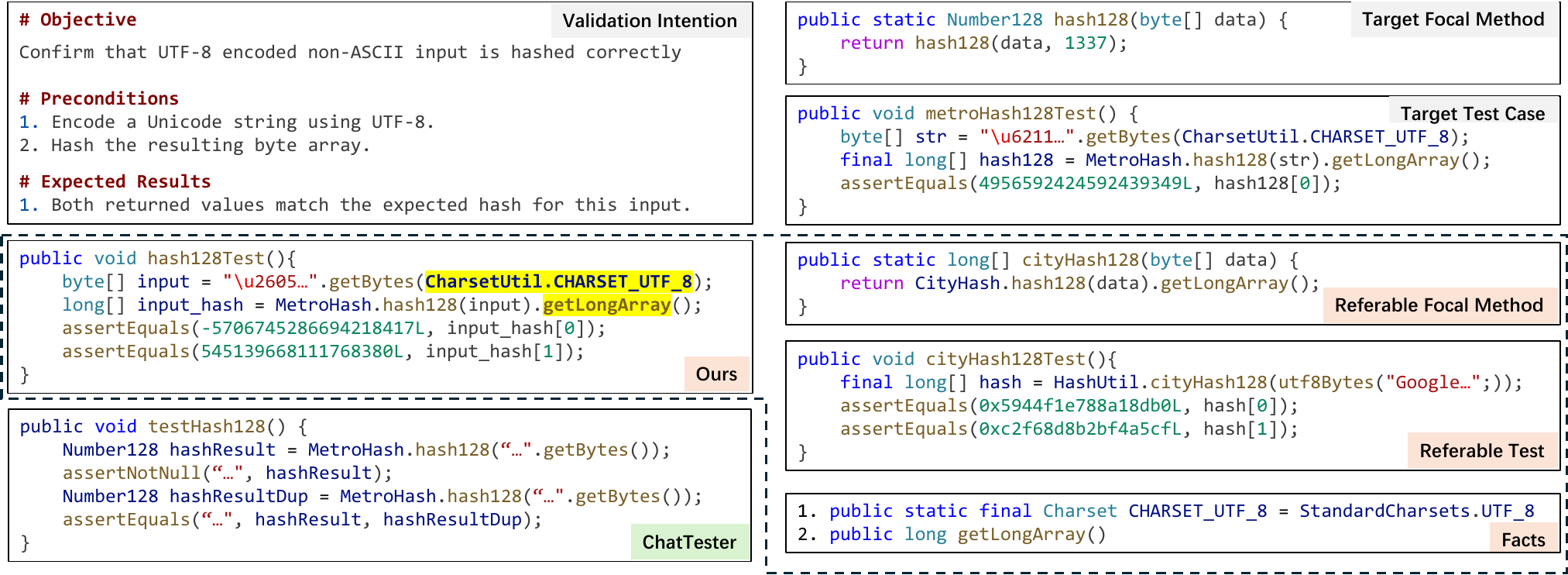}
    \caption{Example comparison of generated tests for the same validation intention: \tool vs. ChatTester. \tool achieves higher alignment with the ground truth (CMS: 100\% vs 8\%).}
    \label{fig:rq1_example}
    \vspace{-12pt}
\end{figure}

\noindent\textbf{Performance on Semantic Relevance}
As shown in \autoref{tab:rq1-summary},
\tool outperforms the baselines regarding line coverage overlap (including exact match and common coverage ratio) and common mutation score.
Specifically, \tool achieves the highest line coverage overlap (87.56\% and 93.40\%) and CMS results (76.35\%), outperforming all state-of-the-art baselines on ``LLM-Inferred'' dataset, e.g., TELPA by 20.41\%, 16.90\%, and 28.09\%, respectively.
The results of ``CMS$_{pair}$'' further show the comparison of \tool against each baseline on their common passing tests.
\tool consistently outperforms all baselines in validation intention alignment---for example, \tool outperforms TELPA by 6.85\% (81.11\% vs. 74.26\%) and 4.22\% (73.63\% vs. 69.41\%) on ``Human-Written'' and ``LLM-Inferred'' dataset, respectively.

As for CodeBLEU, the results of \tool, ChatTester, and TELPA are much smaller than DA's result by about 30\%.
Our results show that DA achieves a significant fit to the distribution of the 12 projects, and also imply the effectiveness of the strategy for mitigating the LLM's memorization effect.
Moreover, we find that many generated tests by \tool are semantically correct but syntactically different from the ground-truth.

\autoref{fig:rq1_example} presents a case where \tool significantly outperforms all baselines.
The intended test scenario is to validate that the focal method \texttt{hash128()} correctly hashes a UTF-8-encoded string.
The tricky parts are twofold: 
(1) \textit{what to test}: recognize that the input should be encoded in UTF-8 and know the correct output; and 
(2) \textit{how to test}: two crucial facts, i.e., \texttt{CHARSET\_UTF\_8} and \texttt{get\-Long\-Array()}, are required for the pre- and post-setup.
Without the intention description and the facts, the generated tests fail their assertions (e.g., DA) or do not align with the validation intention (e.g., ChatTester and EvoSuite).
For example, 
ChatTester uses the platform-default charset via \texttt{get\-Bytes()} (which may not be UTF-8) and employs trivial assertions that do not verify the value returned by \texttt{get\-Long\-Array()} against the correct result, yielding a CMS of only 8\%.
In contrast, \tool retrieves a reference and collects the required facts---both semantically close to the intention and frequently co-occurring with the focal method---achieving a CMS of 100\%.
More details are presented at ~\cite{dtester}.

\noindent\textbf{Performance on Different LLMs.}
The column ``Ours (DS)'' in \autoref{tab:rq1-summary} shows that \tool powered by DeepSeek-V3.2
still outperforms the baselines.
Generally, the retrieved test reference and crucial code facts
provide useful and enriched external knowledge,
making our approach less dependent on the capability of foundation models.
We also observe that \tool(DS) underperforms \tool (O1),
caused by its limited inference performance of 
adopting crucial code facts in editing the referable test.

As a result, by capturing project-specific information and integrating the description of validation intention,
\tool establishes itself as a new state-of-the-art approach for practical projects.

%% file: tables/rq1-summary.tex
\begin{table}[]
\caption{Performance of \tool and baselines (in \%). ``Evo.'' and ``Chat.'' represent EvoSuite and ChatTester, respectively.}%
\vspace{-6pt}
\label{tab:rq1-summary}
\resizebox{\columnwidth}{!}{
\begin{tabular}{llrrrrrrrrrrrr}
\toprule
\multicolumn{2}{c}{\multirow{2}{*}{\textbf{Metric}}} & \multicolumn{6}{c}{\textbf{LLM-Inferred (2536 tests)}}                                                                                                                                                                                 & \multicolumn{6}{c}{\textbf{Human-Written (86 tests)}}                                                                                                                                                                                  \\ \cmidrule(lr){3-8} \cmidrule(lr){9-14}
\multicolumn{2}{c}{}                                 & \multicolumn{1}{c}{\textbf{Evo.}} & \multicolumn{1}{c}{\textbf{Chat.}} & \multicolumn{1}{c}{\textbf{DA}} & \multicolumn{1}{c}{\textbf{TELPA}} & \multicolumn{1}{c}{\textbf{Ours(DS)}} & \multicolumn{1}{c}{\textbf{Ours(O1)}} & \multicolumn{1}{c}{\textbf{Evo.}} & \multicolumn{1}{c}{\textbf{Chat.}} & \multicolumn{1}{c}{\textbf{DA}} & \multicolumn{1}{c}{\textbf{TELPA}} & \multicolumn{1}{c}{\textbf{Ours(DS)}} & \multicolumn{1}{c}{\textbf{Ours(O1)}} \\ \midrule
\multicolumn{2}{l}{Compilation Failure}              & 0.08                                  & 26.17                                   & 35.84                           & 29.44                              &       4.38                            &              3.55                         & 0.00                                  & 41.43                                   & 24.32                           & 29.33                               &     2.60                                 &   3.90                            \\
\multicolumn{2}{l}{Execution Failure}                & 0.00                                  & 4.42                                    & 4.06                            & 4.32                               &       2.80                            &              2.96                         & 0.00                                  & 2.86                                    & 12.16                           & 12.00                               &     5.19                                 &   1.30                            \\
\multicolumn{2}{l}{Assertion Failure}                & 7.87                                  & 1.08                                    & 17.10                           & 2.12                               &       3.11                            &              1.50                         & 15.07                                 & 4.29                                    & 18.92                           & 1.33                                &     3.90                                 &   0.00                            \\
\multicolumn{2}{l}{Successful Pass}                  & 92.04                                 & 68.34                                   & 43.00                           & 64.12                              &       89.72                           &              92.00                         & 84.93                                 & 51.43                                   & 44.59                           & 57.33                              &     88.31                                &   94.81                           \\ \midrule
\multicolumn{2}{l}{Exact Match (on coverage)}        & 42.42                                 & 58.85                                   & 52.31                           & 67.15                              &       87.24                           &              87.56                         & 23.29                                 & 27.78                                   & 63.64                           & 72.09                              &     64.62                                &   70.42                           \\
\multicolumn{2}{l}{Common Coverage Ratio}            & 66.52                                 & 70.27                                   & 69.48                           & 76.50                              &       92.91                           &              93.40                         & 58.16                                 & 47.93                                   & 81.49                           & 85.40                              &     83.72                                &   89.28                           \\ \midrule
\multicolumn{2}{l}{CMS}                              & 30.32                                 & 38.72                                   & 41.82                           & 48.26                              &       55.30                           &              76.35                         & 35.62                                 & 36.25                                   & 42.39                           & 50.10                              &     60.53                                &   85.18                           \\ \midrule
\multirow{2}{*}{CMS$_{pair}$}       & Baseline       & 29.57                                 & 51.94                                   & 52.61                           & 69.41                              &       -                               &              -                         & 36.75                                 & 34.88                                   & 59.93                           & 74.26                                  &     -                                    &   -                               \\
                                    & Ours           & 73.35                                 & 74.53                                   & 73.69                           & 73.63                              &       -                               &              -                         & 85.41                                 & 86.09                                   & 85.87                           & 81.11                                  &     -                                    &   -                               \\ \midrule
\multicolumn{2}{l}{CodeBLEU}                         & 27.16                                 & 45.93                                   & 72.70                           & 55.04                              &       46.87                           &              41.73                         & 29.77                                 & 44.27                                   & 78.80                           & 58.65                              &     50.41                                &   44.42                           \\ \bottomrule
\end{tabular}
}
\vspace{-6pt}
\end{table}

%% file: rq2.tex
\subsection{Granularity of Validation Intention (RQ2)}
\label{sec:test-desc}
\subsubsection{Setup}
Given the effort required to write a validation intention description, real-world projects may impose different levels of detail for it.
We investigate the effectiveness of the description across five levels of detail, including full description (\textit{full}), objective only (\textit{obj}), objective \& preconditions (\textit{obj\&pre}), objective \& expected results (\textit{obj\&exp}), and no description (\textit{none}).

\input{tables/rq2-test-desc-o1}

\begin{figure}[t]
  \centering
  \includegraphics[width=1\textwidth]{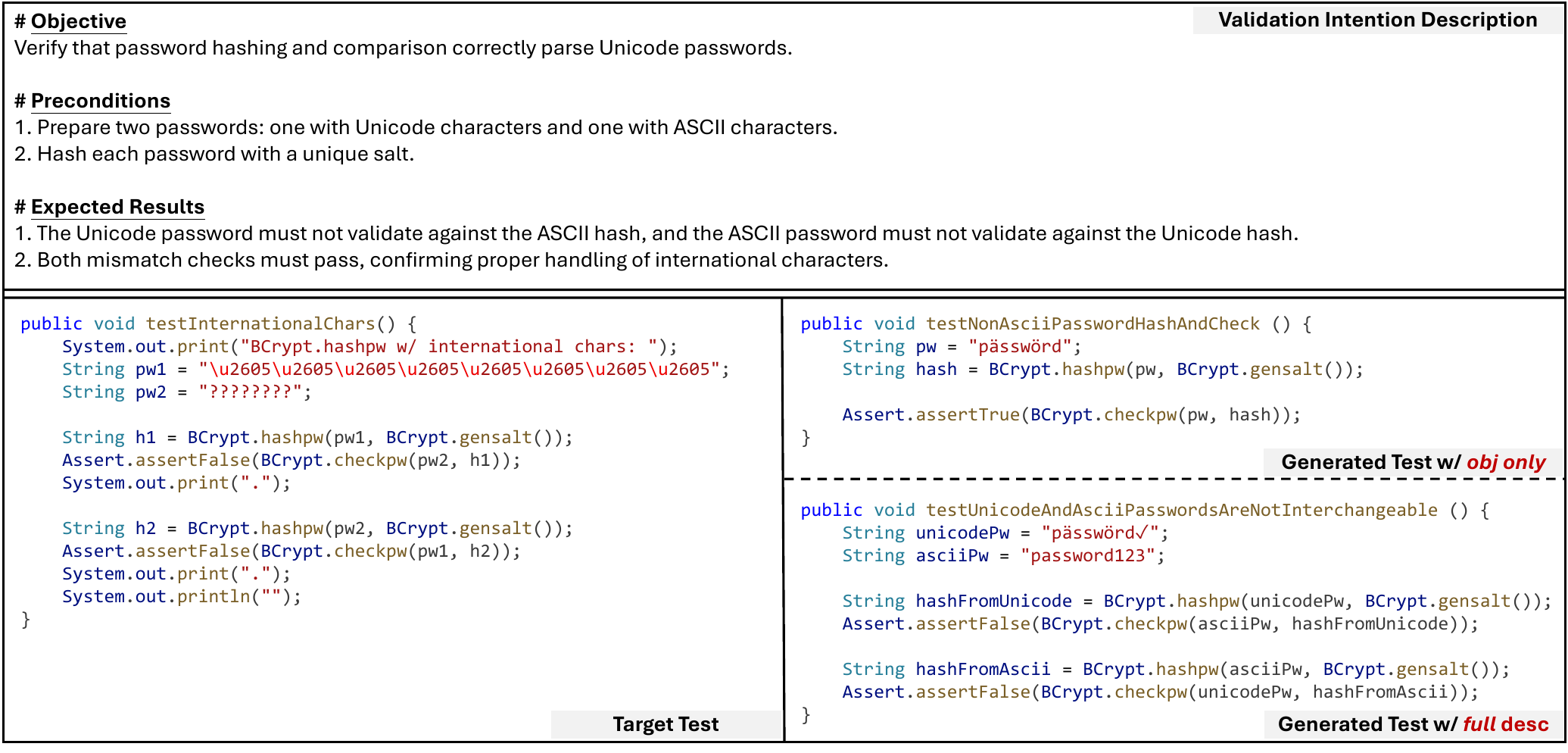}
  \vspace{-12pt}
  \caption{Illustrative example of validation intention granularity: target test \texttt{testInternationalChars} and \tool outputs generated using the full intention description and the objective-only description.}
  \label{fig:desc_example}
  \vspace{-12pt}
\end{figure}

\subsubsection{Results}
As shown in \autoref{tab:rq2-test-desc-o1},
providing any component can improve performance (excluding metrics related to successful pass),
indicating the effectiveness of the three components.
Notably, even providing only a test objective within 50 words can significantly improve LLMs in validation intention alignment, yielding $CMS_{pair}$ improvements of 15.1\% and 27.7\% over \tool without intention description (72.9\% vs 57.7\% and 86.5\% vs 58.8\%) and 4.2\% and 6.9\% over TELPA (73.6\% vs 69.4\% and  81.1\% vs 74.2\%).

This improvement is expected, as even a brief Objective description captures the developer's core intention and helps guide test generation accordingly.
For example, \autoref{fig:desc_example} presents a target test \texttt{testInternationalChars} from the blade project~\cite{blade}, alongside its validation intention description (from ``LLM-Inferred'' dataset) and the corresponding tests generated with the full description and the Objective only, respectively.
The target test primarily verifies correct hashing of non-ASCII passwords.
It also incorporates a personalized intention: validating hashing and comparison correctness for a password consisting of black stars (i.e.,\texttt{u2605}) by cross-checking results with another password, rather than simply comparing plain and hashed values.
In this case, the Objective portion of the generated description---``Verify that password hashing and comparison correctly parse Unicode passwords''---succinctly captures the main intention. Consequently, the test generated with the Objective only already aligns with the main intention, using a password containing non-ASCII characters ``ä'' and ``ö''.
We also observed that the test generated with the full description aligns even more closely,
as the Preconditions and Expected Results provide complementary details reflecting the personalized intention.
In contrast, without such an intention description,
ChatTester, and DA instead generate tests with common ASCII passwords, failing to capture the intended validation goal.
These results highlight the practicality of validation intention descriptions, as even a concise Objective can substantially improve validation intention alignment.

We also analyze suboptimal generated tests to understand their causes.
One common issue arises when the validation intention description is overly general.
For example, a target test \texttt{codePointsAllIncludedRange} from project yavi~\cite{yavi} defines a whitelist by a project-specific API \texttt{CodePoints.Range.of()}, restricting input characters to a–z and A–Z.
Due to our strict constraints on reverse engineering intention descriptions (e.g., minimizing inclusion of program elements),
the derived description simplifies this intention to: ``Define a whitelist which allows only uppercase and lowercase letters.''
As a result, \tool generates a test that implements this intention naively by enumerating uppercase and lowercase letters in a list.
While the generated test executes successfully, it partially misaligns with the target test.
This misalignment is reflected in the CMS metric, as some mutants associated with \texttt{CodePoints.Range.of()} remain surviving.
This result exposes a limitation of \tool---the suboptimal ability of fact discrimination component to capture fine-grained semantic distinctions between codes and intention descriptions.
We attribute this limitation partly to the weaker semantic understanding of the lightweight embedding model used (i.e., CodeT5 Embedding).
Addressing this challenge is left for future work.

Furthermore, even without a validation intention description, \tool still outperforms the state-of-the-art ChatTester---showing, for example, 29.54\% and 35.27\% gains in CMS (see the columns ``ChatTester'' in \autoref{tab:rq1-summary} and the columns ``none'' in \autoref{tab:rq2-test-desc-o1}).
This result underscores the standalone effectiveness of our referable-test retrieval and crucial-facts discrimination methods.
We also observe a small increase in successful passes when descriptions are removed.
This occurs because the generation now faces fewer constraints, making it easier for to generate.
However, this ease comes at the cost of lower CMS, indicating weaker alignment with the intended validation.

More experiments for \tool powered by DeepSeek-V3.2, available at ~\cite{dtester}, also support the above conclusion.

%% file: tables/rq2-test-desc-o1.tex
\begin{table}[t]
\caption{Performance of \tool across five levels of intention description detail (in \%).}
\vspace{-6pt}
\label{tab:rq2-test-desc-o1}
\centering
\resizebox{0.95\columnwidth}{!}{
\begin{tabular}{lcrrrrrrrrrr}
\toprule
\multicolumn{2}{c}{\multirow{2}{*}{\textbf{Metric}}} & \multicolumn{5}{c}{\textbf{LLM-Inferred}}                                                                                                                    & \multicolumn{5}{c}{\textbf{Human-Written}}                                                                                                                          \\ \cmidrule(lr){3-7} \cmidrule(lr){8-12}
\multicolumn{2}{c}{}                                 & \multicolumn{1}{c}{\textbf{full}} & \multicolumn{1}{c}{\textbf{obj}} & \multicolumn{1}{c}{\textbf{obj\&pre}} & \multicolumn{1}{c}{\textbf{obj\&exp}} & \multicolumn{1}{c}{\textbf{none}} & \multicolumn{1}{c}{\textbf{full}} & \multicolumn{1}{c}{\textbf{obj}} & \multicolumn{1}{c}{\textbf{obj\&pre}} & \multicolumn{1}{c}{\textbf{obj\&exp}} & \multicolumn{1}{c}{\textbf{none}} \\ \midrule
\multicolumn{2}{l}{Compilation Failure}       & 3.55    & 3.62   & 3.03      & 3.39      & 2.84   & 3.90     & 1.32    & 1.30       & 3.90       & 0.00    \\
\multicolumn{2}{l}{Execution Failure}         & 2.96    & 2.88   & 2.80      & 2.88      & 2.99   & 1.30     & 2.63    & 0.00       & 2.60       & 0.00    \\
\multicolumn{2}{l}{Assertion Failure}         & 1.50    & 1.81   & 1.65      & 2.46      & 1.38   & 0.00     & 1.32    & 3.90       & 3.90       & 1.30    \\
\multicolumn{2}{l}{Successful Pass}           & 92.00   & 91.69  & 92.52     & 91.27     & 92.79  & 94.81    & 94.74   & 94.81      & 89.61      & 98.70   \\ \midrule
\multicolumn{2}{l}{Exact Match (on coverage)} & 87.56   & 87.41  & 87.75     & 87.11     & 71.53  & 70.42    & 75.71   & 68.57      & 72.06      & 57.53   \\
\multicolumn{2}{l}{Common Coverage Ratio}    & 93.40   & 93.38  & 93.55     & 93.31     & 84.54  & 89.28    & 88.91   & 87.18      & 87.83      & 75.59   \\ \midrule
\multicolumn{2}{l}{CMS}                       & 76.35   & 70.49  & 72.03     & 71.47     & 68.26  & 85.18    & 74.15   & 73.79      & 73.80      & 71.52   \\ \midrule
\multirow{2}{*}{CMS$_{pair}$}    & Variant    & -       & 73.02  & 74.07     & 71.96     & 57.79  & -        & 76.11   & 73.63      & 74.35      & 58.77   \\
                                 & Full       & -       & 72.52  & 73.29     & 72.56     & 72.93  & -        & 85.52   & 85.77      & 85.56      & 86.49   \\ \midrule
\multicolumn{2}{l}{CodeBLEU}                  & 41.73   & 41.83  & 41.62     & 41.63     & 41.53  & 44.42    & 47.60   & 45.84      & 44.45      & 41.96  \\ \bottomrule

\end{tabular}
}
\end{table}

%% file: rq3.tex
\subsection{Ablation Study (RQ3)}\label{sec:retrieval-contribution}
\input{tables/rq3-ablation-o1}

\subsubsection{Setup}
To assess the contribution of the retrieval module, we remove referable tests from the prompts.
Regarding the discriminator's contribution to overall performance,  we ablate it by removing facts from the prompt.

\subsubsection{Results}
As reported in \autoref{tab:rq3-ablation-o1},
the column ``\textit{No Ref}'' presents the performance results when the retrieval function is ablated.
Compared to \tool, the test generation performance declines sharply.
For example, 
the performance in CMS drops by 2.45\% and 11.00\% on ``LLM-Inferred'' and ``Human-Written'' datasets, respectively.
Similar degradations occur in all other metrics.
The results indicate that referable tests can help reveal implicit project-specific test idioms and test scenarios that the LLM alone cannot infer.

We also evaluated the robustness of \tool by analyzing its performance across a spectrum of similarity bins (i.e., $[0.2, 0.4]$, $(0.4, 0.6]$, $(0.6, 0.8]$, and $(0.8, 1.0]$) calculated using the Levenshtein-based edit similarity between the target and referable tests. 
We observed that, while higher similarity correlates with improved generation quality, \tool maintains robust performance even in low-similarity regimes, consistently outperforming the \textit{``No Ref''} variant. 
For example, in the Successful Pass metric, \tool achieved $92.68\%$, $95.83\%$, $97.06\%$, and $96.15\%$, respectively. 
This suggests that 
(1) even low-similarity references could provide useful structural scaffolds and project-specific idioms and 
(2) LLM is relatively robust to irrelevant information.

As for the contribution of crucial facts, 
the column ``\textit{No Fact}''
shows that removing facts from prompts likewise impairs performance, e.g., 3.91\% and 6.38\% of degradation in CMS on the two datasets. 
This confirms the effectiveness of the discriminator.
Experiments for \tool powered by DeepSeek-V3.2, available at ~\cite{dtester}, also support the above conclusion.

%% file: tables/rq3-ablation-o1.tex
\begin{table}[]
\caption{Ablation study results for \tool (in \%).}
\vspace{-6pt}
\label{tab:rq3-ablation-o1}
\centering
\resizebox{0.7\columnwidth}{!}{
\begin{tabular}{llrrrrrr}
\toprule
\multicolumn{2}{c}{\multirow{2}{*}{\textbf{Metric}}} & \multicolumn{3}{c}{\textbf{LLM-Inferred}}                                                               & \multicolumn{3}{c}{\textbf{Human-Written}}                                                                     \\ \cmidrule(lr){3-5}  \cmidrule(lr){6-8} 
\multicolumn{2}{c}{}                                 & \multicolumn{1}{c}{\textbf{Full}} & \multicolumn{1}{c}{\textbf{No Fact}} & \multicolumn{1}{c}{\textbf{No Ref}} & \multicolumn{1}{c}{\textbf{Full}} & \multicolumn{1}{c}{\textbf{No Fact}} & \multicolumn{1}{c}{\textbf{No Ref}} \\ \midrule
\multicolumn{2}{l}{Compilation Failure}       & 3.55           & 4.41            & 5.76           & 3.90     & 2.60       & 7.79      \\
\multicolumn{2}{l}{Execution Failure}         & 2.96           & 3.55            & 5.64           & 1.30     & 1.30       & 2.60      \\
\multicolumn{2}{l}{Assertion Failure}         & 1.50           & 1.30            & 1.18           & 0.00     & 5.19       & 3.90      \\
\multicolumn{2}{l}{Successful Pass}           & 92.00          & 90.74           & 87.41          & 94.81    & 90.91      & 85.71     \\ \midrule
\multicolumn{2}{l}{Exact Match (on coverage)} & 87.56          & 87.65           & 86.53          & 70.42    & 69.57      & 68.75     \\
\multicolumn{2}{l}{Common Coverage Ratio}     & 93.40          & 93.61           & 92.84          & 89.28    & 87.43      & 84.08     \\ \midrule
\multicolumn{2}{l}{CMS}                       & 76.35          & 72.44           & 73.90          & 85.18    & 78.80      & 74.18     \\ \midrule
\multirow{2}{*}{$CMS_{pair}$}   & Ablation   & -              & 72.65           & 68.31          & -        & 73.23      & 65.85     \\
                                 & Full       & -              & 72.94           & 72.72          & -        & 84.72      & 84.47     \\ \midrule
\multicolumn{2}{l}{CodeBLEU}                  & 41.73          & 41.19           & 34.73          & 44.42    & 45.12      & 35.14     \\ \bottomrule

\end{tabular}
}
\vspace{-12pt}
\end{table}

%% file: 6_discussion.tex
\section{Discussion}
\subsection{Limitations and Future Work}

\subsubsection{Quality of Validation Intention Descriptions} In this work, validation intention descriptions generated by LLMs or written by developers are matched with ground-truth tests, filtering out low-quality and misaligned intentions.
We attempted to make them as natural as possible---applying various constraints on generation and performing manual checks---yet a concern remains: human-written descriptions can be incomplete and ambiguous.
Determining when a description is low quality and guiding users to craft higher-quality descriptions remains open work for future research.

\subsubsection{Generalizability on Newly Established Repositories}
In this paper, we target well-established repositories with sufficient human-written tests.
The performance of \tool might not generalize to newly established repositories.
Although we evaluated \tool on repositories with relatively few tests (e.g., 49 for \textit{spark}), its effectiveness on repositories whose test suites are both sparse and low-quality is still unknown.
Quantifying \tool's usefulness in early-stage projects constitutes important future work.

\subsection{Threats to Validity}
\label{sec:threats}

The first threat involves the synthesized validation intention descriptions. 
Due to the limited availability of human-written validation intentions in open-source projects, we use LLMs to generate validation intention descriptions for existing tests. 
However, these synthesized descriptions may differ from those that developers would write. 
To mitigate this, we curate a set of human-written validation intention descriptions as a complementary dataset.

The second threat involves potential data leakage from LLMs, a common issue also faced by related work~\cite{chattester}.
The tests in our evaluation dataset could be part of training data in LLMs, which might lead to overestimation of LLM's capability in test generation.
To mitigate this threat, we design a prompt that instructs LLMs to ``unlearn'' the memorized information from the specified repository.
Experimental observations confirm its effectiveness in preventing the generation of program elements (e.g., method invocations) that do not appear in the input context.

Finally, a threat arises from the nondeterministic nature of LLMs. To alleviate this, we set the temperature parameter to zero, constraining randomness and ensuring the most deterministic response possible~\cite{tosem_determinism_gpt}.

%% file: 7_related_work.tex
\section{Related Work} \label{sec:related-work}

\paragraph{Coverage-based software testing}

Software testing is traditionally considered a constraint-solving problem to generate tests to cover targeted program branches.
Typical test generation solution includes dynamic and static symbolic execution \cite{cadar2008klee, braione2017combining, braione2018sushi, godefroid2005dart, sen2005cute} and search-based software testing \cite{fraser2011evosuite, arcuri2008search, braione2017combining, godoy2021enabledness, lin2021graph, lemieux2023codamosa, pacheco2007randoop, lin2020recovering}.
While test coverage is an important metric for test completeness, readability and quality are also crucial for practical tests~\cite{readability-1,readability-2,quality-1,quality-2}.
Moreover, many tests are knowledge-driven.
Thus, we design \tool, an LLM-based solution. \tool uses the general programming knowledge within the LLM and enhances the LLM with project-specific knowledge, which is complementary to coverage-based test generators.

\paragraph{LLM-based test generation}
With the emergence of LLMs, researchers have made significant advances in code generation~\cite{roziere2023code}.
Considering test code is a special form of code, researchers have leveraged LLMs in software testing~\cite{li2023nuances, schafer2023empirical,xia2024fuzz4all, alshahwan2024automated,generator_yanjie,gao2025promptalchemistautomatedllmtailored,mezzaro2024empirical,qi2021dreamloc,qi2026generalizing}, considering the test generation problem as a translation problem from focal method to test code \cite{nie2023learning, tufano2020unit, kang2023large, dinella2022toga,chattester,IntUT,issta24_test_adaption,wen2025variable}.
One relevant work is ChatTester \cite{chattester}, which designs effective test-generating prompts and validates the tests based on a program analyzer and compiler.
Different from ChatTester, \tool is a retrieval-and-edit solution.
We further develop our contribution to how to effectively edit a referable test by discriminating the crucial project-specific information. 

Additionally, IntUT~\cite{IntUT} leverages generated test intentions to guide LLMs. A test intention in IntUT targets a specific branch within the focal method, specifying input parameters and the expected output when that branch is exercised. Consequently, IntUT is designed to maximize code coverage. Furthermore, expected outputs are derived directly from the focal method, implying that the focal method must return constant values.
In contrast, our work introduces validation intentions, which are written by developers to reflect requirement fulfillment rather than coverage. Also, our approach imposes no constraints on the return behavior of the focal method.

%% file: 8_conclusion.tex
\section{Conclusion} 

In this work, we propose \tool, which can generate project-specific tests with validation intention.
Our solution is motivated by the empirical observation that mature software projects possess abundant code assets that can guide the generation of new tests.
\tool captures the project-specific knowledge by retrieving referable tests and collecting crucial facts, and derives practical tests using LLMs.
We extensively evaluate \tool on 3,680 tests from 12 projects.
The results show that, compared to the state-of-the-art approaches, \tool generates far more practical tests with better alignment to developers' validation intention.

\section{Data Availability}
Our source code and experimental data are available at \cite{dtester}.